\documentclass[a4paper,superscriptaddress,aps,11pt]{article}

\usepackage{graphicx,rotating,hyperref,slashed,amsmath,xcolor,amssymb,amsfonts,colortbl,cite,subcaption}
\usepackage[totalheight = 23cm, totalwidth = 17cm]{geometry}
\usepackage{simplewick}
\usepackage{axodraw2}
\usepackage{ulem}

\definecolor{bluc}{cmyk}{1,1,0,0.1}
\definecolor{rossoCP3}{cmyk}{0,.88,.77,.40}
\definecolor{rosso}{cmyk}{0,1,1,0.4}
\definecolor{rossos}{cmyk}{0,1,1,0.55}
\definecolor{rossoc}{cmyk}{0,1,1,0.2}
\definecolor{verdes}{cmyk}{0.92,0,0.59,0.4}

\hypersetup{colorlinks, bookmarksopen, bookmarksnumbered,
citecolor=verdes, linkcolor=bluc, pdfstartview=FitH, urlcolor=rossos}

\newcommand{\calC}{{\cal C}}

\newcommand{\calL}{{\cal L}}
\newcommand{\calO}{{\cal O}}
\newcommand{\calP}{{\cal P}}
\newcommand{\calR}{{\cal R}}

\newcommand{\calT}{{\cal T}}
\newcommand{\mpl}{m_{\rm Pl}}

\renewcommand*{\thefootnote}{\fnsymbol{footnote}}

\begin{document}

\begin{titlepage}

\rightline{\footnotesize{APCTP-Pre2024-020}}

\begin{center}

\vskip 3em

{\Large \bf 
Scalar one-loop tensor power spectrum during single-field inflation
}

\vskip 3em

{\large Jiwon Kong$^{a,}$\footnote{\label{ref:footnote1}Equal contribution},
Jieun Jeon$^{a,}$\textsuperscript{\ref{ref:footnote1}}
and 
Jinn-Ouk Gong$^{a,b,}$\footnote{Corresponding author}}

\vskip 0.5cm

{\it
$^{a}$Department of Science Education, Ewha Womans University, Seoul 03760, Korea
\\
$^{b}$Asia Pacific Center for Theoretical Physics, Pohang 37673, Korea
}

\end{center}

\vskip 1.2cm

\begin{abstract}

We calculate the scalar-induced one-loop correction to the power spectrum of tensor perturbations produced during single-field slow-roll inflation. We find that the correction is given by the square of the product of the slow-roll parameter and the tree-level scalar power spectrum. We also discuss the implications of the logarithmic contribution.

\end{abstract}

\end{titlepage}

\renewcommand*{\thefootnote}{\arabic{footnote}}
\setcounter{footnote}{0}

\newpage

\section{Introduction}
\label{sec:intro}

The observations of the cosmic microwave background (CMB) over last decades have confirmed that its temperature is almost the same everywhere. Because the size of the causal patch when the CMB was generated was much smaller than what corresponds to the current observable universe, the homogeneous CMB as observed raises the question how the temperature in each causally disconnected patch could be nearly the same. Further, there exist tiny temperature fluctuations of the CMB of $\calO(10^{-5})$, correlated beyond each patch, with the correlation strength between two points, i.e. power spectrum, more or less of the same amplitude no matter how far they are separated~\cite{Planck:2018vyg}. Unless one resorts to extremely finely tuned initial conditions of the big bang, a certain physical mechanism is required that can explain both the homogeneity and isotropy of the CMB and small perturbations with almost scale-invariant power spectrum.

The most promising candidate of such a mechanism is cosmic inflation~\cite{Guth:1980zm,Linde:1981mu,Albrecht:1982wi}, usually assumed to be driven by a scalar field called the inflaton. During inflation, the universe experiences accelerated expansion so that each causal patch is driven exponentially to spatial homogeneity and isotropy. Furthermore, tiny quantum fluctuations are stretched to become classical perturbations on super-horizon scales~\cite{Mukhanov:1981xt}. These primordial perturbations can be translated into those in the spatial metric: One corresponding to the perturbation in the spatial curvature, and the others to the traceless and traceless components identified as the gravitational waves (GWs), usually referred to respectively as scalar and tensor perturbations. We can easily compute the properties of these perturbations under the assumption that the inflaton rolls down a flat enough potential so that the acceleration of the inflaton can be ignored~\cite{Mukhanov:2005sc,Weinberg:2008zzc,Baumann:2022mni}. With this ``slow-roll'' approximation, the power spectrum of the scalar perturbations and that of the tensor perturbations are nearly scale-invariant.

The prediction of these scale-invariant spectra is general and can be realized in many models of inflation~\cite{Martin:2013tda}. On one hand, this shows the robustness of the framework of inflation. But on the other hand, it implies that we cannot distinguish different inflation models, and the physics relevant during inflation would remain elusive. We need other observable targets beyond the power spectra to learn the physics of the early universe. Fortunately, we do have: Gravity is of non-linear nature and thus provides non-linear interactions beyond quadratic one responsible for the robust tree-level power spectra. These non-linear interactions give rise to higher-order correlation functions, such as the bispectrum and trispectrum, which are prime targets of the observation programs in the coming decades~\cite{Giannantonio:2011ya,Karagiannis:2018jdt,Sohn:2019rlq}. At the same time, they also mean non-linear contributions to the power spectra -- quantum loop corrections~\cite{Weinberg:2005vy,Sloth:2006az,Sloth:2006nu,Seery:2007we,Seery:2007wf,Dimastrogiovanni:2008af,Adshead:2008gk,Adshead:2009cb,Senatore:2009cf,Bartolo:2010bu}.


In this article, we compute the scalar-induced one-loop corrections to the power spectrum of tensor perturbations in single-field slow-roll inflation. While most works on one-loop corrections are regarding the scalar power spectrum, we explicitly compute the tensor power spectrum with the one-loop corrections induced by scalar curvature perturbation\footnote{
Note that although the one-loop corrections to the tensor power spectrum are calculated in~\cite{Tan:2019czo} previously, the loops are induced not by the scalar curvature perturbation, but by other matter contents such as a spectator scalar field.
}. In principle, such loop corrections may be large enough to be observationally relevant, enabling us to distinguish different inflation models and hence different microscopic physics behind each model. If so, unlike the tree level calculations, loop corrections may well be dependent on the whole history of the universe, not just on the behavior around the moment of horizon exit. Even if the magnitude of the one-loop corrections turns out too small to be verifiable observationally, by explicit computations of the loop corrections we can find unexpected subtleties and gain new insights into the theory. Moreover, we can constrain otherwise unobservable hidden sector via loop corrections~\cite{delRio:2018vrj} and consistency of effective field theories on curved space-time~\cite{Melville:2021lst}.

On top of these general motivations for studying loop corrections, there is another specific motivation for our study of scalar one-loop corrections to the tensor power spectrum. After the discovery of GWs by LIGO~\cite{LIGOScientific:2016aoc}, primordial black holes (PBHs) have been considered as a possible source of the merger events of binary black holes as well as a candidate of dark matter~\cite{Carr:2020gox,Carr:2020xqk,Escriva:2022duf}. PBHs are formed when a region enters the horizon with the density perturbation above a certain threshold value~\cite{Harada:2013epa}. Such a large amplitude of perturbation could be originated from the preceding epoch of inflation, e.g. when the inflaton rolls over an extremely flat potential, $dV/d\phi \approx 0$. This phase is called ultra slow-roll (USR)~\cite{Tsamis:2003px,Kinney:2005vj}, and the scalar perturbation experiences exponential growth during USR~\cite{Germani:2017bcs,Motohashi:2017kbs,Ballesteros:2017fsr}. Such a large amplitude of scalar perturbation can result in large amplitude of tensor perturbations sourced by scalar perturbations -- large induced gravitational waves during inflation. Thus, on very general ground, it is theoretically interesting to study scalar one-loop corrections to the tensor power spectrum.


\section{Structure of scalar-induced one-loops}
\label{sec:structure}

We consider the Einstein gravity with a minimally coupled canonical scalar field $\phi$, which we identify as the inflaton:
\begin{equation}
\label{eq:model}
S = \int d^4x \sqrt{-g} \bigg[ \frac{\mpl^2}{2}R - \frac{1}{2}g^{\mu\nu}\partial_\mu\phi\partial_\nu\phi - V(\phi) \bigg] \, ,
\end{equation}
where $g = \det{g_{\mu\nu}}$ is the determinant of the metric tensor, $\mpl^2 \equiv 1/(16\pi G)$ is the reduced Planck mass, $R$ is the Ricci scalar and $V(\phi)$ is the potential of the inflaton $\phi$. We adopt the Arnowitt-Deser-Misner convention of the metric~\cite{Arnowitt:1962hi}:
\begin{equation}
ds^2 = -N^2dt^2 + \gamma_{ij} ( \beta^idt + dx^i ) (\beta^jdt + dx^j ) \, .
\end{equation}
The spatial metric $\gamma_{ij}$ includes both scalar and tensor perturbations, denoted respectively by $\calR$ and $h_{ij}$\footnote{
The perturbations $\calR$ and $h_{ij}$ defined in this form are non-linearly conserved on super-horizon scales~\cite{Salopek:1990jq,Maldacena:2002vr,Lyth:2004gb}. One may introduce the scalar and tensor perturbations into the metric using different parametrization, for example $\gamma_{ij} = a^2(t) \big( e^{2\calC} \big)_{ij}$ or $\gamma_{ij} = a^2(t) \big( \delta_{ij} + 2\calC_{ij} \big)$, according to which the correlation functions become different at non-linear order. The (in)variance of the correlation functions, especially for scalar-tensor mixed ones, depending on the parametrization of cosmological perturbations -- that is, under field redefinition -- is a subtle issue but an extensive study on it is beyond the scope of the present work.
}:
\begin{equation}
\label{eq:spatial-metric}
\gamma_{ij} = a^2(t) e^{2\calR} \big( e^h \big)_{ij} \, ,
\end{equation}
with $h^i{}_{j,i} = h^i{}_i = 0$, i.e. $h_{ij}$ is transverse and traceless. In agreement with the homogeneous Friedmann-Robertson-Walker metric $ds^2 = -dt^2 + a^2(t)\delta_{ij}dx^idx^j$, the lapse function $N$ and the shift vector $\beta_i$ can be set including perturbations as:
\begin{align}
N & = 1 + \alpha \, ,
\\
\beta_i & = \psi_{,i} + N_i \, ,
\end{align}
with $N^i{}_{,i} = 0$. Further, we also decompose $\phi$ into the background and perturbation:
\begin{equation}
\phi = \phi_0 + \delta\phi \, .
\end{equation}

\begin{figure}
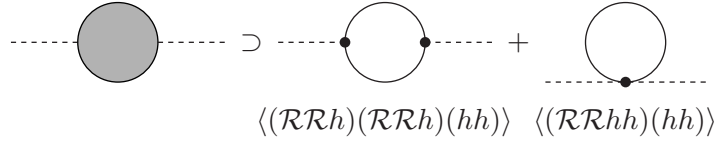

\begin{center}
 \begin{axopicture}(260,50)
  \DashLine(0,30)(25,30){2}
  \GCirc(40,30){15}{0.7}
  \DashLine(55,30)(80,30){2}
  \Text(90,30)[]{$\supset$}
  \DashLine(100,30)(125,30){2}
  \Vertex(125,30){2}
  \Arc(140,30)(15,0,360)
  \Vertex(155,30){2}
  \DashLine(155,30)(180,30){2}
  \Text(140,0)[]{$\langle(\calR\calR h)(\calR\calR h)(hh)\rangle$}
  \Text(190,30)[]{$+$}
  \DashLine(200,15)(260,15){2}
  \Vertex(230,15){2}
  \Arc(230,30)(15,0,360)
  \Text(230,0)[]{$\langle(\calR\calR hh)(hh)\rangle$}
 \end{axopicture}
\end{center}
\caption{Connected diagrams relevant for scalar one-loop corrections to the tensor power spectrum. Solid and dashed lines denote respectively scalar and tensor propagators, and dots represent interaction vertices. The first diagram includes two three-point interaction vertices between two scalar and one tensor perturbations, and the second diagram includes one four-point interaction vertex between two scalar and two tensor perturbations, respectively. We also have tadpole-like diagrams, which do not contribute to the tensor power spectrum.}
\label{fig:diagrams}
\end{figure}

Given the model by \eqref{eq:model}, now we consider the contributions relevant for the scalar-induced one-loop corrections to the tensor power spectrum. Diagramatically, the connected scalar one-loop diagrams for the tensor power spectrum are shown in Figure~\ref{fig:diagrams}. They include three- and four-point interaction vertices between the scalar and tensor perturbations. Thus, to compute the scalar one-loop corrections to the tensor power spectrum, we need up to the quartic-order action for scalar-tensor interactions. We can obtain the action up to quartic-order in perturbations straightly by expanding the action \eqref{eq:model} up to fourth order in perturbations of our interest $\calR$ and $h_{ij}$, with the solutions of $\alpha$, $\psi$ and $N_i$ given in terms of $\calR$ and $h_{ij}$ being plugged back into the action. Note that in some previous studies (see e.g.~\cite{Senatore:2009cf}) the interactions are based on effective theory approach, but here we calculate up to the desired quartic-order action from the Einstein-Hilbert term.

Then, as illustrated in Appendix~\ref{app:Hamiltonian}, we can find the full Hamiltonian by the Legendre transformation of the Lagrangian and, in the interaction picture (see e.g.~\cite{Wang:2013zva}) the scalar-tensor interaction Hamiltonian leading-order in slow-roll relevant for the diagrams given in Figure~\ref{fig:diagrams} are found to be:
\begin{align}
\label{eq:H3}
H_{(sst)} 
& = 
- a \epsilon \mpl^2 \int d^3x \calR_{,i} \calR_{,j} h^{ij}
\, ,
\\
\label{eq:H4}
H_{(sstt)} 
& = 
a^3 \frac{\mpl^2}{2} \int d^3x \Bigg\{ \frac{1}{4} \alpha_2^{(ss)} \dot{h}^{ij}\dot{h}_{ij}
+ \frac{1}{4a^2} \alpha_2^{(ss)} h^{ij,k}h_{ij,k}- \frac{1}{a^2} \Bigg[
\frac{1}{2} \Big( \beta_{2i,j}^{(ss)} - \beta_{2j,i}^{(ss)} \Big) \dot{h}^{ik}h_k{}^j 
- \frac{1}{2} \beta_2^{k(ss)} \dot{h}^{ij}h_{ij,k} \Bigg] 
\nonumber\\
& \hspace{7em}
+ 2\epsilon\dot\calR^2 \alpha_2^{(tt)} + \frac{\epsilon}{a^2} \Big(
\calR_{,i}\calR_{,j}h^{ik}h_k{}^j + 2\calR^{,i}\calR_{,i}\alpha_2^{(tt)} 
+ 4\dot\calR\calR^{,i}\beta_{2i}^{(tt)} \Big) \Bigg\}
\, ,
\end{align}
where, with $H \equiv \dot{a}/a$ being the Hubble parameter, $\epsilon \equiv -\dot{H}/H^2$ is the slow-roll parameter and is assumed to be very small during slow-roll inflation, the numbers in the subscript of $\alpha$ and $\beta_i$ denote the order in perturbations, and the superscripts $(ss)$ and $(tt)$ mean the solutions are given respectively by the products of two scalar and two tensor perturbations. These solutions are explicitly given later.

\section{Calculations of loops}
\label{sec:calculation}

\subsection{Loop with two three-point vertices}

We first begin with the diagram for the loop with two three-point interaction vertices. The possible schematic contraction between perturbation fields is of the form:
\begin{equation}
\begin{split}
\begin{axopicture}(80,30)
 \DashLine(0,15)(25,15){2}
 \Vertex(25,15){2}
 \Arc(40,15)(15,0,360)
 \Vertex(55,15){2}
 \DashLine(55,15)(80,15){2}
\end{axopicture}
\end{split}
\quad \mapsto \quad
\langle(
\contraction{}{\calR}{\calR h)(}{\calR}
\contraction[2ex]{\calR}{\calR}{h)(\calR}{\calR}
\contraction[3ex]{\calR\calR}{h}{)(\calR\calR h)(}{h}
\contraction[1ex]{\calR\calR h)(\calR\calR}{h}{)(h}{h}
\calR\calR h)(\calR\calR h)(hh)\rangle
\, .
\end{equation}
Then using the in-in formalism~\cite{Schwinger:1960qe,Keldysh:1964ud} the leading non-zero contribution from this diagram is given by the term including two cubic-order interaction Hamiltonian in the Dyson series:
\begin{align}
\label{eq:RRh}
\big\langle h_{ij}(\pmb{k}_1) h_{kl}(\pmb{k}_2) \big\rangle
=
\Re \bigg\{ \calT
\int_{\tau_0}^\tau d\tau_1 \int_{\tau_0}^\tau d\tau_2 \Big\langle
& 
H_{(sst)}(\tau_1) h_{ij}(\pmb{k}_1) h_{kl}(\pmb{k}_2) H_{(sst)}(\tau_2) 
\nonumber\\
& 
- h_{ij}(\pmb{k}_1) h_{kl}(\pmb{k}_2) H_{(sst)}(\tau_1) H_{(sst)}(\tau_2) \Big\rangle \bigg\}
\, ,
\end{align}
where $\calT$ denotes time-ordering and $d\tau \equiv dt/a$ is the conformal time.

\begin{figure}
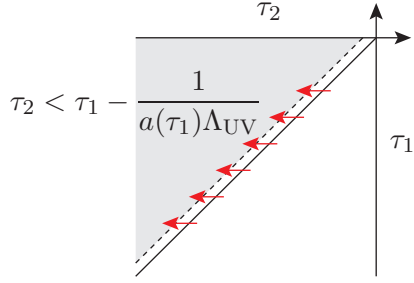

\begin{center}
 \begin{axopicture}(100,100)
  {\SetColor{LightGray}
  \FTri(0,5)(0,90)(85,90)}
  \LongArrow(0,90)(100,90)
  \LongArrow(90,0)(90,100)
  \Text(50,100)[]{$\tau_2$}
  \Text(100,50)[]{$\tau_1$}
  \Line(0,0)(90,90)
  \DashLine(0,5)(85,90){2}
  \Text(0,65)[]{$\tau_2 < \tau_1 - \dfrac{1}{a(\tau_1)\Lambda_\text{UV}}$}
  {\SetColor{Red}
  \LongArrow(73,70)(63,70)
  \LongArrow(63,60)(53,60)
  \LongArrow(53,50)(43,50)
  \LongArrow(43,40)(33,40)
  \LongArrow(33,30)(23,30)
  \LongArrow(23,20)(13,20)}
 \end{axopicture}
\end{center}
\caption{If $\tau_1>\tau_2$, the time-ordered integrals $\calT \int_{\tau_0}^\tau d\tau_1 \int_{\tau_0}^\tau d\tau_2$ can be written as $\int_{\tau_0}^\tau d\tau_1 \int_{\tau_0}^{\tau_1} d\tau_2$. Reflecting the ultraviolet cutoff $a(\tau_1)\Lambda_\text{UV}$, we also remove from the time integral the domain where the time interval is shorter than what corresponds to $a(\tau_1)\Lambda_\text{UV}$. The domain to be integrated is denoted by gray shade.}
\label{fig:t-integral}
\end{figure}

The Bunch-Davies mode function solutions for the scalar and tensor perturbations denoted respectively by $\calR_k(\tau)$ and $h_k^{(\lambda)}(\tau)$ in perfect de Sitter limit, where $H$ is constant and $a(\tau) = -1/(H\tau)$, are given by
\begin{align}
\calR_k(\tau) 
& =
\frac{1}{\sqrt{4\epsilon}} \frac{iH}{k^{3/2}\mpl} (1+ik\tau) e^{-ik\tau}
\, ,
\\
\label{eq:modefct}
h_k^{(\lambda)}(\tau)
& =
\frac{iH}{k^{3/2}\mpl} (1+ik\tau) e^{-ik\tau}
\, .
\end{align}
Here, $\lambda$ represents two independent polarizations of the tensor perturbations. Note that we can include slow-roll corrections to the above de Sitter solutions and proceed the calculations described below~\cite{delRio:2018vrj}, which however demands more tedious works. Plugging these solutions into \eqref{eq:RRh} and performing naively the integrals, we can find that the result is ultraviolet divergent as $q_\text{UV}^3/k^3$, where $k \equiv |\pmb{k}_1| = |\pmb{k}_2|$ and $q_\text{UV}$ is the naive ultraviolet cutoff for the internal momentum of the loop. Thus, to obtain meaningful result, we need to regularize this ultraviolet divergence by introducing an appropriate comoving cutoff as $a\Lambda_\text{UV}$~\cite{Senatore:2009cf}, where the scale factor $a$ should be dependent on the earlier interaction time, say, $\tau_2$ if $\tau_1 > \tau_2$: 
\begin{equation}
\label{eq:momentum-cutoff}
\int_0^\infty dq ~ \to ~ \int_0^{a(\tau_2)\Lambda_\text{UV}} dq \, .
\end{equation}
This means when the interaction is first encountered at $\tau_2$, the modes with higher momenta than the cutoff do not contribute. Note that since the momentum cutoff is time-dependent, we first integrate the internal momentum and then perform the time integrals. Accordingly, $\tau_2$ can be pushed up to $\tau_1$, off the same cutoff as depicted in Figure~\ref{fig:t-integral}. This corresponds to removing in the time integral the domain where the time interval is shorter than the same cutoff, $1/(a\Lambda_\text{UV})$:
\begin{equation}
\label{eq:modified-range}
\int_{\tau_0}^\tau d\tau_1 \int_{\tau_0}^{\tau_1} d\tau_2 ~ \to ~
\int_{\tau_0}^\tau d\tau_1 \int_{\tau_0}^{\tau_1-\frac{1}{a(\tau_1)\Lambda_\text{UV}}} d\tau_2
\end{equation}

Once we have set up the ultraviolet cutoff to regulate divergences, it is straightforward to proceed the calculations. We relegate the details to Appendix~\ref{app:Rhh-calculation}, and present only the final result here. Performing the integrals, we obtain the one-loop correction \eqref{eq:RRh} evaluated at $\tau$:
\begin{align}
\label{eq:RRh-result}
\big\langle h_{ij}(\pmb{k}_1) h_{kl}(\pmb{k}_2) \big\rangle
& = 
(2\pi)^3 \delta^{(3)}(\pmb{k}_{12}) \frac{2\pi^2}{k^3}
\sum_{\lambda,\lambda'} \frac{e_{ij}^{(\lambda)}(\hat{\pmb{k}}_1)e_{kl}^{(\lambda')}(\hat{\pmb{k}}_2)}{128\pi^2} 
\frac{H^2}{\mpl^2} \frac{8}{\mpl^2} \bigg( \frac{H}{2\pi} \bigg)^2 
\nonumber\\
& \hspace{11em}
\times
\frac{1}{135} \Big\{ -133 + 240 \big[ \gamma + \log2 + \log(-k\tau) \big] \Big\}
\nonumber\\
& \quad
+ \text{terms dependent on the power of } \Lambda_\text{UV} \, ,
\end{align}
where $\gamma \approx 0.577216$ is the Euler-Mascheroni constant. Note that we face no infrared divergence here. Since the terms dependent on the power of $\Lambda_\text{UV}$ can be removed by adding appropriate counter terms, such terms do not contribute to the one-loop correction. Also, the slow-roll approximation we are adopting remains valid for $\exp(-1/\xi) \ll -k\tau \ll \exp(1/\xi)$ for some small number $\xi$ that could be comparable to the typical value of the slow-roll parameter~\cite{Gong:2001he}. Thus we can evaluate the logarithm at any convenient time around horizon crossing, making this contribution negligible.

\subsection{Loop with a four-point vertex}

Next, we consider the diagram for the loop with a four-point interaction vertex. The possible contraction is of the following form:
\begin{equation}
\label{eq:H4contraction}
\begin{split}
\begin{axopicture}(60,30)
 \DashLine(0,0)(60,0){2}
 \Vertex(30,0){2}     
 \Arc(30,15)(15,0,360)
\end{axopicture}
\end{split}
\quad \mapsto \quad
\langle(
\contraction{}{\calR}{}{\calR}
\contraction[1ex]{\calR\calR}{h}{h)(}{h}
\contraction[2ex]{\calR\calR h}{h}{)(h}{h}
\calR\calR hh)(hh)\rangle
\, ,
\end{equation}
and the leading non-zero contribution is given by
\begin{align}
\label{eq:RRhh}
\big\langle h_{ij}(\pmb{k}_1) h_{kl}(\pmb{k}_2) \big\rangle
=
2 \Im \bigg[ 
\int_{\tau_0}^\tau d\tau' \Big\langle
h_{ij}(\pmb{k}_1) h_{kl}(\pmb{k}_2) H_{(sstt)}(\tau') \Big\rangle \bigg]
\, .
\end{align}
To proceed the calculations, as can be read from \eqref{eq:H4}, we need the solutions for the lapse and shift perturbations up to second-order. The second-order solutions can be found by solving algebraically the constraint equations~\cite{Arroja:2008ga}, and to leading order in slow-roll they are:
\begin{align}
\alpha_2^{(ss)}
& =
\frac{\epsilon}{H} \Delta^{-1} 
\Big( \dot\calR\calR_{,i} \Big)^{,i}
\, ,
\\
\alpha_2^{(tt)}
& =
\frac{1}{8H} \Delta^{-1} 
\Big( \dot{h}^{ij}\Delta h_{ij} + \dot{h}^{ij,k}h_{ij,k} \Big)
\, ,
\\
N_{2i}^{(ss)}
& =
4a^2\epsilon \Delta^{-1} \Big[ \Delta^{-1} \Big( \dot\calR\calR_{,j} \Big)^{,j}{}_{,i} - \dot\calR\calR_{,i} \Big]
\, ,
\\
N_{2i}^{(tt)}
& =
\frac{a^2}{2} \Delta^{-1} \Big[ \Delta^{-1} \Big( \dot{h}^{jk}\Delta h_{jk} + \dot{h}^{jk,l}h_{jk,l} \Big)_{,i}
+ \dot{h}^{jk}h_{ij,k} - \dot{h}^{jk}h_{jk,i} - h^{ij}\dot{h}_{ik,j} \Big]
\, ,
\\
\psi_2^{(ss)}
& =
- \frac{\epsilon}{2H}a^2 \Delta^{-1} \bigg[ 6H\Delta^{-1} \Big( \dot\calR\calR_{,i} \Big)^{,i} + \dot\calR^2 + \frac{1}{a^2}\calR^{,i}\calR_{,i} \bigg]
\, ,
\\
\psi_2^{(tt)}
& =
- \frac{a^2}{16H} \Delta^{-1} \bigg[ 6H \Delta^{-1} \Big( \dot{h}^{ij}\Delta h_{ij} + \dot{h}^{ij,k}h_{ij,k} \Big) 
+ \dot{h}^{ij}\dot{h}_{ij} + \frac{1}{a^2}h^{ij,k}h_{ij,k} \bigg]
\, ,
\end{align}
where $\Delta^{-1}$ is the inverse Laplacian operator. We see that the solutions given by the product of two scalar perturbations are suppressed by the slow-roll parameter $\epsilon$, so $H_{(sstt)}$ given by \eqref{eq:H4} is $\calO(\epsilon)$.

As we can see, all the second-order solutions of the lapse and shift include inverse Laplacian operators. In the momentum space, such terms contribute $\delta^{(3)}(\pmb{q}_{12})/q_{12}^2$ for the scalar solutions because, as we can see from the schematic contraction \eqref{eq:H4contraction}, two $\calR$'s in $H_{(sstt)}$ should be contracted to each other, and $\delta^{(3)}(\pmb{k}_{12})/k_{12}^2$ for the tensor solutions because two $h$'s in $H_{(sstt)}$ should be contracted to the external legs. Both diverge in the infrared regime, so we may introduce an infrared cutoff $\Lambda_\text{IR}$ and as usual subtract such infrared-divergent contributions, $\langle(\calR\calR hh)(hh)\rangle - \langle(\calR\calR hh)(hh)\rangle_\text{IR}$. Note that in principle we can and should elaborate the infrared contribution as much as the ultraviolet one as we did in the previous section, and properly considering the infrared regime would bring subtle issues, such as time-dependent renormalized coefficients~\cite{Huenupi:2024ksc}.
Then the only remaining term in the quartic interaction Hamiltonian is $\calR_{,i}\calR_{,j}h^{ik}h_k{}^j$, and using \eqref{eq:RRhh} we can straightly compute the contribution from this term, with the details being given in Appendix~\ref{app:RRhh-calculation}, and find that it is proportional to $\Lambda_\text{UV}^2/H^2$, the same as what we can face for a scalar field in flat space-time. Thus, it can be absorbed by counter terms and in the end there remains no meaningful contribution in the one-loop correction \eqref{eq:RRhh}.

\subsection{Total one-loop corrections}

From the results we obtained in the previous sections, we see that the relevant scalar-induced one-loop correction to the tensor power spectrum is \eqref{eq:RRh}. To compare with the tree-level tensor power spectrum $\calP_h = 8[H/(2\pi)]^2/\mpl^2$, it is more convenient to write \eqref{eq:RRh} in terms of $\calP_h$. Or, since the loop is mediated by the scalar propagators, we can use the tree-level scalar power spectrum $\calP_\calR = H^2/(8\pi^2\epsilon\mpl^2) = \calP_h/(16\epsilon)$ to write the total scalar-induced one-loop correction:
\begin{align}
\label{eq:1-loop-final}
\delta\calP_h
& = 
(\epsilon\calP_\calR)^2 \frac{1}{135} \Big\{ -133 + 240 \big[ \gamma + \log2 + \log(-k\tau) \big] \Big\}
\nonumber\\
& 
\sim \calO(1) \times (\epsilon\calP_\calR)^2  
\nonumber\\
&
\sim \calO(10^{-2}) \times \calP_h^2 
\, ,
\end{align}
where for definiteness we have set the logarithm zero. Equivalently, we can write \eqref{eq:1-loop-final} in terms of the ratio of $H$ to $\mpl$:
\begin{equation}
\delta\calP_h \sim \calO(10^{-4}) \times \bigg( \frac{H}{\mpl} \bigg)^4
\, .
\end{equation}

\section{Discussions}
\label{sec:discuss}

In this article, we have studied the scalar-induced one-loop corrections to the tensor power spectrum during single-field slow-roll inflation. The result is consistent with our expectations: Since the relevant one-loop correction \eqref{eq:RRh} includes two tree-level scalar power spectra and two three-point interaction vertices of strength $\calO(\epsilon)$ each, the contribution of the one-loop corrections to the power spectrum is $\big( \epsilon\calP_\calR \big)^2 \sim \calP_h^2$. This is equivalent to $(H/\mpl)^4$.

An interesting question regarding general loop corrections is the divergent behavior. Usually, the cutoff scale of the internal momentum enters into the loop correction logarithmically in standard quantum field theory, and one may well expect a similar dependence. Meanwhile, here we have found the logarithmic correction as $\log(-k\tau) = \log[k/(aH)]$ (see also~\cite{Maity:2023qzw}). This well-known logarithmic term arises in e.g. higher-order corrections in slow-roll approximation~\cite{Gong:2001he}, and is usually interpreted as a late-time super-horizon effect. At first look, this does not seem consistent with our expectation. Nevertheless, we can relate this to the ultraviolet cutoff in a few different ways:
\begin{enumerate}

\item As $k\to0$, this logarithmic contribution diverges. That is, the modes with comoving momentum $k$ much smaller than the comoving Hubble horizon $1/(aH)$ pile up on super-horizon scale and these redshifted, infrared modes leads to large effects. This divergence, however, can be under control once we realize that the inflationary patch of our observational interest is not infinitely extended spatially so the momentum faces a cutoff at the scale corresponding to the inverse of the finite size of the patch $L$. Denoting this finite size as $L^{-1} \sim a\Lambda_\text{UV}$, this logarithm reads $\log(-k\tau) \sim \log[L^{-1}/(aH)] \sim \log(\Lambda_\text{UV}/H)$. This interpretation is along the arguments of some literature on loop corrections, for example~\cite{Seery:2007we,Seery:2007wf,Dimastrogiovanni:2008af}.

\item For a fixed value of $k$, we face infinity from this logarithm as $\tau\to0$. But we know that $\tau = 0$ corresponds to infinite future in perfect de Sitter space. Thus precisely speaking, evaluating the correlation function at $\tau = 0$ does not make sense since inflation never ends, or it is eternal, and there will be no observer to measure any correlation functions. For our calculations to make sense on a certain mode, inflation must end at some finite time and we therefore cut off the evaluation time well before future infinity. This evaluation time may well be after the moment of horizon crossing, long enough for the mode to become a constant value. Indeed, if we choose $\tau \sim -[a(\tau)\Lambda_\text{UV}]^{-1}$ in the time integral of $\tau_1$ in \eqref{eq:modified-range}, then the logarithm reads $\log(-k\tau) \sim \log[a H/(a\Lambda_\text{UV})] \sim \log(\Lambda_\text{UV}/H)$.

\item We may not have to take a particular limit where the logarithm diverges. Instead, we can think of the argument of the logarithm in terms of the physical momentum in such a way that $-k\tau = (k/a)/H$. This is essentially the ratio of the physical momentum $k/a$ probed by the power spectrum to the energy scale during inflation $H$. The physical momentum can be as large as the ultraviolet cutoff, $k/a \lesssim \Lambda_\text{UV}$. Thus the logarithm reads at the ultraviolet cutoff $\log(-k\tau) \sim \log(\Lambda_\text{UV}/H)$.

\end{enumerate}
Thus, although not explicit, we can read the expected logarithmic dependence of the ultraviolet cutoff $\log(\Lambda_\text{UV}/H)$.

Finally, as mentioned in Introduction, we expect during USR the amplitude of the scalar perturbation experiences exponential enhancement large enough to produce PBHs after inflation. Since $\epsilon \propto a^{-6}$ during USR, we find another slow-roll parameter $\eta \equiv \dot\epsilon/(H\epsilon) = -6$ which is also assumed to be very small during standard slow-roll inflation. Thus, as done in this article, simply working with the least slow-roll suppressed terms may not capture the features during USR~\cite{Maity:2023qzw,Kristiano:2022maq,Firouzjahi:2023btw,Ballesteros:2024zdp}. Further, the boundary terms -- especially temporal ones -- may well be included to correctly account for the final result. Nevertheless we believe our study of the scalar-induced one-loop corrections to the tensor power spectrum during slow-roll inflation is the starting point, revealing subtle conceptual and practical aspects. We hope to return to these points soon.

\subsection*{Acknowledgements}

We thank Perseas Christodoulidis for discussions at early stages of this work.
We are also indebted to Jason Kristiano, Andrew Long, Gonzalo Palma, Chang Sub Shin, Takahiro Tanaka, Masahide Yamaguchi and Yuhang Zhou for helpful discussions.
We also thank the anonymous editor for her/his constructive feedback that improved the contents of this article.
This work is supported in part by the Ministry of Education of the Republic of Korea and the National Research Foundation of Korea (2024S1A5C3A03046593) and by Basic Science Research Program through the National Research Foundation of Korea (RS-2024-00336507). JG also acknowledges the Ewha Womans University Research Grant of 2024 (1-2024-0651-001-1).
JG is grateful to the Asia Pacific Center for Theoretical Physics for hospitality while this work was under progress.

\appendix

\renewcommand{\theequation}{\thesection.\arabic{equation}}

\section{Higher-order Hamiltonian in the interaction picture}
\label{app:Hamiltonian}
\setcounter{equation}{0}

In this section, we sketch briefly the derivation of the interaction Hamiltonian up to quartic order, the leading contributions of which in slow-roll are presented in \eqref{eq:H3} and \eqref{eq:H4} in the main text. Our starting point is the action \eqref{eq:model} expanded up to fourth order in perturbations of our interest $\calR$ and $h_{ij}$ defined in the spatial metric as in \eqref{eq:spatial-metric}. In doing so, we first work in the flat gauge where $\calR = 0$ and the scalar degree of freedom is $\delta\phi$. The reason why we choose the flat gauge, although eventually we are interested in $\calR$, is because naively expanding the action from the beginning in the comoving gauge, where we set $\delta\phi = 0$, gives us terms that are not slow-roll suppressed at all. Since $\calR$ in perfect de Sitter case $\epsilon=0$ is a pure gauge mode, the action for $\calR$ should disappear in that limit. Another reason is that the variable that can be canonically quantized at non-linear order is not $\calR$ but (say) $\calR_n$, defined by
\begin{equation}
\calR_n \equiv -\frac{H}{\dot\phi_0}\delta\phi \sim \frac{\delta\phi}{\sqrt{\epsilon}} \, ,
\end{equation}
so that in terms of $\delta\phi$ the slow-roll suppression of the curvature perturbation becomes clear. Note that $\calR_n$ is related to the non-linearly conserved curvature perturbation $\calR$ by
\begin{equation}
\calR = \calR_n + \frac{\eta}{4}\calR_n^2 + \frac{1}{H}\dot\calR_n\calR + \cdots \, .
\end{equation}

Once the action expanded up to quartic order in $\delta\phi$ and $h_{ij}$ -- with the solutions for $\alpha$ and $\beta_i$ being plugged back -- the {\it full} non-linear conjugate momentum of $\delta\phi$ is, to leading order in slow-roll, 
\begin{equation}
\label{eq:Pi-phi}
\Pi_{\delta\phi} \equiv \frac{\delta\calL}{\delta\dot{\delta\phi}} 
= a^3 \Bigg( \dot{\delta\phi} - \frac{1}{a^2} \psi_{1,i} \delta\phi^{,i}
- \frac{\dot\phi_0}{2\mpl^2H} \dot{\delta\phi}\delta\phi + \frac{\dot\phi_0^3}{4\mpl^4H^2} \delta\phi^2 \Bigg)
+ \delta\calL_4^{(\delta\phi)}
\, ,
\end{equation}
where $\psi_1 \equiv -a^2 \epsilon \big( H/\dot\phi_0 \Delta^{-1}\delta\phi \big)^\cdot$ is the linear longitudinal solution for the shift vector in terms of $\delta\phi$, and the last term $\delta\calL_4^{(\delta\phi)}$ denotes the contributions coming from the quartic order Lagrangian, which thus are cubic order in $\delta\phi$. Similarly, we find the full non-linear conjugate momentum of $h_{ij}$ as, also to leading order in slow-roll,
\begin{equation}
\label{eq:Pi-h}
\Pi_{ij} \equiv \frac{\delta\calL}{\delta\dot{h}^{ij}}
= \frac{a^3\mpl^2}{4} \Bigg( \dot{h}_{ij} + \frac{\dot\phi_0}{a^2\mpl^2H}\delta\phi\psi_{1,ij}
- \frac{\dot\phi_0}{2\mpl^2H} \delta\phi\dot{h}_{ij} - \frac{1}{a^2} h_{ij,k}\psi_1{}^{,k} \Bigg)
+ \delta\calL_4^{(h)}
\, .
\end{equation}
Replacing $\dot{\delta\phi}$ and $\dot{h}_{ij}$ with $\Pi_{\delta\phi}$ and $\Pi_{ij}$ respectively in the Lagrangian and perform the Legendre transformation gives the Hamiltonian. Then we can easily find that both $\delta\calL_4^{(\delta\phi)}$ and $\delta\calL_4^{(h)}$ cancel away, and the terms in the parentheses in \eqref{eq:Pi-phi} and \eqref{eq:Pi-h}, which come from the cubic order Lagrangian, contribute additionally to the quartic order interaction Hamiltonian. Note that for the pure tensor sector there is no time derivative of ${h}_{ij}$ at cubic order so that there is no new contribution at quartic order from the conjugate momentum to Hamiltonian.

Finally, to compute the higher-order correlation functions, we transit to the interaction picture by replacing the conjugate momenta in the Hamiltonian with those from the quadratic parts:
\begin{align}
\Pi_{\delta\phi} & \to \Pi_{\delta\phi} = a^3 \dot{\delta\phi} \, ,
\\
\Pi_{ij} & \to \Pi_{ij} = \frac{a^3\mpl^2}{4} \dot{h}_{ij} \, .
\end{align}
Then we obtain the cubic and quartic order interaction Hamiltonian from which we can compute three- and four-point correlation functions, rediscovering the well-known result that $H_3 = -L_3$. Note that the new contributions to quartic order are, as can be read from \eqref{eq:Pi-phi} and \eqref{eq:Pi-h}, all slow-roll suppressed as $\calO(\epsilon^2)$ or even higher, and thus are sub-leading compared to the terms shown in \eqref{eq:H4}, which are of slow-roll order $\calO(\epsilon)$.

\section{Detail on the loop with two three-point vertices}
\label{app:Rhh-calculation}
\setcounter{equation}{0}

In this section, we present more details of the calculations of \eqref{eq:RRh} that give \eqref{eq:RRh-result}. In terms of the Fourier modes, the leading cubic-order interaction Hamiltonian is written as:
\begin{equation}
H_{(sst)} 
=
a\epsilon\mpl^2 \int \frac{d^3q_1d^3q_2d^3q_3}{(2\pi)^{3\cdot2}} \delta^{(3)}(-\pmb{q}_{123})
(q_{2i}q_{3j}) h^{ij}(\pmb{q}_1) \calR(\pmb{q}_2) \calR(\pmb{q}_3) 
\, ,
\end{equation}
where the time dependence of the Fourier modes is suppressed. Denoting $h_k^{(\lambda)}(\tau) \equiv u_k(\tau)$ for both polarization indices, then $\calR_k(\tau) = u_k(\tau)/\sqrt{4\epsilon}$, and we find:
\begin{align}
\label{eq:int-HOH}
&
\Big\langle H_{(sst)}(\tau_1) h_{ij}(\pmb{k}_1) h_{kl}(\pmb{k}_2) H_{(sst)}(\tau_2) \Big\rangle
\nonumber\\
& = 
(2\pi)^3 \delta^{(3)}(\pmb{k}_{12}) \frac{\mpl^4}{4H^4} \frac{1}{\tau_1^2\tau_2^2}
\sum_{\lambda,\lambda'} e_{ij}^{(\lambda)}(\hat{\pmb{k}}_1) e_{kl}^{(\lambda')}(\hat{\pmb{k}}_2)
\nonumber\\
& \quad
\times 
\int \frac{d^3q_1d^3q_2}{(2\pi)^3} \delta^{(3)}(\pmb{k}-\pmb{q}_{12}) \big|u_k(\tau)\big|^2
\big[ u_k(\tau_1) u_{q_1}(\tau_1) u_{q_2}(\tau_1) \big]
\big[ u_k(\tau_2) u_{q_1}(\tau_2) u_{q_2}(\tau_2) \big]^* 
\nonumber\\
& \quad
\times 
\sum_{s,s'} (q_{1a}q_{2b}) e^{ab}_{(s)}(-\hat{\pmb{k}}_1)
(q_{1c}q_{2d}) e^{cd}_{(s')}(-\hat{\pmb{k}}_2)
\, ,
\\
\label{eq:int-OHH}
&
\Big\langle h_{ij}(\pmb{k}_1) h_{kl}(\pmb{k}_2) H_{(sst)}(\tau_1) H_{(sst)}(\tau_2) \Big\rangle
\nonumber\\
& =
(2\pi)^3 \delta^{(3)}(\pmb{k}_{12}) \frac{\mpl^4}{4H^4} \frac{1}{\tau_1^2\tau_2^2}
\sum_{\lambda,\lambda'} e_{ij}^{(\lambda)}(\hat{\pmb{k}}_1) e_{kl}^{(\lambda')}(\hat{\pmb{k}}_2)
\nonumber\\
& \quad
\times 
\int \frac{d^3q_1d^3q_2}{(2\pi)^3} \delta^{(3)}(\pmb{k}-\pmb{q}_{12}) \big[u_k(\tau)\big]^2
\big[ u_k^*(\tau_1) u_{q_1}(\tau_1) u_{q_2}(\tau_1) \big]
\big[ u_k(\tau_2) u_{q_1}(\tau_2) u_{q_2}(\tau_2) \big]^* 
\nonumber\\
& \quad
\times \sum_{s,s'} (q_{1a}q_{2b}) e^{ab}_{(s)}(-\hat{\pmb{k}}_1)
(q_{1c}q_{2d}) e^{cd}_{(s')}(-\hat{\pmb{k}}_2)
\, .
\end{align}
where $\pmb{k} \equiv \pmb{k}_1 = -\pmb{k}_2$ due to the momentum conservation.

First, using the delta function for the internal momenta we can perform the integration for $\pmb{q}_2$, which simply amounts to writing $\pmb{q}_2 = \pmb{k}-\pmb{q}_1 \equiv \pmb{k}-\pmb{q}$. Then, from the transverse property of the polarization tensor $k^ie_{ij}^{(s)}(\hat{\pmb{k}}) = 0$, we have:
\begin{equation}
(q_{1a}q_{2b}) e^{ab}_{(\lambda)}(-\hat{\pmb{k}})
(q_{1c}q_{2d}) e^{cd}_{(\lambda')}(\hat{\pmb{k}})
~ \to ~
(q_aq_b) e^{ab}_{(\lambda)}(-\hat{\pmb{k}}) (q_cq_d) e^{cd}_{(\lambda')}(\hat{\pmb{k}})
\, .
\end{equation}
To proceed the momentum integrations further, we set up the momenta as follows. Thanks to the rotational symmetry, without losing generality we align $\pmb{k}$ along the $z$-axis, and $\pmb{q}$ has the polar angle $\theta$ and is confined in the $xz$-plane, i.e. the azimuthal angle $\phi$ is zero.  That is, each vector has the following components:
\begin{align}
\hat{\pmb{k}} & = (0,0,1) \, ,
\\
\pmb{q} & = (q\sin\theta, 0, q\cos\theta) \, ,
\end{align}
and the corresponding polarization tensors for $\hat{\pmb{k}}$ are:
\begin{align}
e_{ij}^{(+)}(\hat{\pmb{k}}) & = \begin{pmatrix} 1 & 0 & 0 \\ 0 & -1 & 0 \\ 0 & 0 & 0 \end{pmatrix} \, ,
\\
e_{ij}^{(\times)}(\hat{\pmb{k}}) & = \begin{pmatrix} 0 & 1 & 0 \\ 1 & 0 & 0 \\ 0 & 0 & 0 \end{pmatrix} \, .
\end{align}
We can then compute explicitly to find\footnote{
If we do not confine $\pmb{q}$ in the $xz$ plane but keep generic $\phi$, we have
\begin{equation*}
\sum_\lambda q_aq_b e^{ab}_{(\lambda)}(\pm\hat{\pmb{k}}) 
= 
q^2(1-\cos^2\theta) \big[ \cos(2\phi) \pm \sin(2\phi) \big] 
\, .
\end{equation*}
}:
\begin{equation}
\sum_\lambda q_aq_b e^{ab}_{(\lambda)}(-\hat{\pmb{k}})
=
\sum_{\lambda'} q_cq_d e^{cd}_{(\lambda')}(\hat{\pmb{k}})
=
q^2(1-\cos^2\theta) \, .
\end{equation}
Note that we find the same results with circular polarization tensors, $e_{ij}^{(s)} \equiv \left(e_{ij}^{(+)}+ise_{ij}^{(\times)}\right)/\sqrt{2}$, with $s = +1$ (right-circular polarization) or $s = -1$ (left-circular one). At this stage, the part of \eqref{eq:int-HOH} dependent on the momentum integrals is
\begin{align}
\label{eq:HOH-momentumint}
\Big\langle H_{(sst)}(\tau_1) h_{ij}(\pmb{k}_1) h_{kl}(\pmb{k}_2) H_{(sst)}(\tau_2) \Big\rangle
&
\supset
\big|u_k(\tau)\big|^2 u_k(\tau_1) u_k^*(\tau_2) 
\nonumber\\
& \quad 
\times
\int \frac{d^3q}{(2\pi)^3} q^4 (1-\cos^2\theta)^2 
u_q(\tau_1) u_{|\pmb{k}-\pmb{q}|}(\tau_1) u_q^*(\tau_2) u_{|\pmb{k}-\pmb{q}|}^*(\tau_2)
\, ,
\end{align}
and similarly for \eqref{eq:int-OHH}.

Next, we introduce new variables:
\begin{align}
t & \equiv \frac{|\pmb{k}-\pmb{q}|+q}{k} \, ,
\\
s & \equiv \frac{|\pmb{k}-\pmb{q}|-q}{k} \, ,
\end{align}
with $1\leq t\leq \infty$ and $-1\leq s\leq 1$. Then, using the explicit form of $u_k(\tau)$ given by \eqref{eq:modefct}, the momentum integral of \eqref{eq:HOH-momentumint} becomes:
\begin{equation}
\frac{k}{(2\pi)^2} \frac{H^4}{2\mpl^4} \int_1^\infty dt \int_{-1}^1 ds 
\bigg[ \frac{(1-s^2) (t^2-1)}{t^2-s^2} \bigg]^2 
\bigg[ 1 + ik\tau_1t - \frac{1}{4} k^2\tau_1^2 \big( t^2-s^2 \big) \bigg] e^{-ik\tau_1t}
\bigg[ 1 - ik\tau_2t - \frac{1}{4} k^2\tau_2^2 \big( t^2-s^2 \big) \bigg] e^{ik\tau_2t}
\, .
\end{equation}
Now we can perform the integrations explicitly with respect to $s$ and $t$. Note that the cutoff in $q$ given by \eqref{eq:momentum-cutoff} corresponds to modifying the integration range of $t$ as:
\begin{equation}
\int_1^\infty dt ~ \to ~ \int_1^{-2\frac{\Lambda_\text{UV}/H}{k\tau_1}} dt \, .
\end{equation}
Performing the integration with respect to $t$ gives the exponential factor $e^{-ik(\tau_1-\tau_2)}$ at the boundary $t=1$ and $\exp\big[2i(\Lambda_\text{UV}/H)(\tau_1-\tau_2)/\tau_2\big]$ at $t=-2\Lambda_\text{UV}/H/(k\tau_1)$. Here, $\tau_2$ can be close to $\tau_1$ by the offset $-1/[a(\tau_1)\Lambda_\text{UV}] = \tau_1/(\Lambda_\text{UV}/H)$, the minimum difference allowed by the cutoff. Thus, compared to $e^{-ik(\tau_1-\tau_2)}$ from $t=1$, the contribution from the boundary $t=-2\Lambda_\text{UV}/H/(k\tau_1)$ is always exponentially rapidly oscillating and we thus only take into account the contribution from the boundary $t=1$. Then it is straightforward to perform the integration with respect to $\tau_2$ and then $\tau_1$. Finally, taking the limit $\Lambda_\text{UV}/H \gg 1$ and $\tau\to0$, we find:
\begin{align}
\label{eq:HOH-result}
&
\int_{\tau_0}^\tau d\tau_1 \int_{\tau_0}^{\tau_1-\frac{1}{a(\tau_1)\Lambda_\text{UV}}} d\tau_2
\Big\langle H_{(sst)}(\tau_1) h_{ij}(\pmb{k}_1) h_{kl}(\pmb{k}_2) H_{(sst)}(\tau_2) \Big\rangle
\nonumber\\
& =
(2\pi)^3 \delta^{(3)}(\pmb{k}_{12}) \frac{2\pi^2}{k^3} 
\sum_{\lambda,\lambda'} \frac{e^{(\lambda)}_{ij}(\hat{\pmb{k}}_1)e^{(\lambda')}_{kl}(\hat{\pmb{k}}_2)}{128\pi^2}
\frac{8}{\mpl^2} \bigg( \frac{H}{2\pi} \bigg)^2 \frac{1}{k^2} 
\bigg( - \frac{4}{3\tau^2} + \frac{4}{3}k^2 + \cdots \bigg)
\, ,
\end{align}
where the ellipses mean the terms imaginary and/or dependent on the power of $\Lambda_\text{UV}/H$, which do not contribute to the final results as we only need real terms and we can remove power-law dependent terms by adding appropriate counter terms. Similarly, we find:
\begin{align}
\label{eq:OHH-result}
&
\int_{\tau_0}^\tau d\tau_1 \int_{\tau_0}^{\tau_1-\frac{1}{a(\tau_1)\Lambda_\text{UV}}} d\tau_2
\Big\langle h_{ij}(\pmb{k}_1) h_{kl}(\pmb{k}_2) H_{(sst)}(\tau_1) H_{(sst)}(\tau_2) \Big\rangle
\nonumber\\
& =
(2\pi)^3 \delta^{(3)}(\pmb{k}_{12}) \frac{2\pi^2}{k^3} 
\sum_{\lambda,\lambda'} \frac{e^{(\lambda)}_{ij}(\hat{\pmb{k}}_1)e^{(\lambda')}_{kl}(\hat{\pmb{k}}_2)}{128\pi^2}
\frac{8}{\mpl^2} \bigg( \frac{H}{2\pi} \bigg)^2 \frac{1}{k^2}
\nonumber\\
& \quad
\times
\bigg\{ - \frac{4}{3\tau^2} - \frac{1}{135}k^2 \big[ - 313 + 240\gamma + 240\log(-k\tau) \big] 
+ \cdots \bigg\}
\, .
\end{align}
Subtracting \eqref{eq:OHH-result} from \eqref{eq:HOH-result} gives our result in the main text, \eqref{eq:RRh-result}.

\section{Detail on the loop with a four-point vertex}
\label{app:RRhh-calculation}
\setcounter{equation}{0}

Here, we give the detailed calculations of \eqref{eq:RRhh} with the interaction Hamiltonian $\calR_{,i}\calR_{,j}h^{ik}h_k{}^j$ that can eventually be removed by appropriate counter terms. The relevant Hamiltonian is:
\begin{align}
H_{(sstt)} 
& = 
\frac{1}{2} a\epsilon\mpl^2 \int d^3x \calR_{,i}\calR_{,j}h^{ik}h_k{}^j
\nonumber\\
& =
\frac{1}{2} a\epsilon\mpl^2 \int \frac{d^3q_1\cdots d^3q_4}{(2\pi)^{3\cdot3}} \delta^{(3)}(-\pmb{q}_{1234})
\calR(\pmb{q}_1)\calR(\pmb{q}_2 h_{(\lambda_1)}(\pmb{q}_3) h_{(\lambda_2)}(\pmb{q}_4)
\sum_{s,s'} (-q_{1i}q_{2j}) 
e^i{}_k^{(s)}(\hat{\pmb{q}}_3) e^{kj}_{(s')}(\hat{\pmb{q}}_4)
\, .
\end{align}
Then we find:
\begin{align}
\label{eq:sstt-int1}
\Big\langle h_{ij}(\pmb{k}_1) h_{kl}(\pmb{k}_2) H_{(sstt)}(\tau') \Big\rangle
& =
(2\pi)^3 \delta^{(3)}(\pmb{k}_{12})  \frac{\mpl^2}{8H^2} \frac{1}{{\tau'}^2}
\sum_{\lambda,\lambda'} 
e_{ij}^{(\lambda)}(\hat{\pmb{k}}_1) e_{kl}^{(\lambda')}(\hat{\pmb{k}}_2)
\nonumber\\
& \quad
\times
\big[ u_k(\tau) \big]^2
\int \frac{d^3q}{(2\pi)^3} \big|u_q(\tau')\big|^2 \big[ u_k^*(\tau') \big]^2
\sum_{s,s'} (q_{i}q_{j}) 
e^i{}_k^{(s)}(-\hat{\pmb{k}}_1) e^{kj}_{(s')}(-\hat{\pmb{k}}_2)
\, .
\end{align}
Again, we align $\pmb{k} = \pmb{k}_1$ along the $z$-axis and confine $\pmb{q}$ in the $xz$-plane. Then, irrespective of the choice of the polarization tensors, we obtain:
\begin{equation}
\sum_{s,s'} (q_{i}q_{j}) 
e^i{}_k^{(s)}(-\hat{\pmb{k}}) e^{kj}_{(s')}(\hat{\pmb{k}})
=
2q^2(1-\cos^2\theta) 
\, .
\end{equation}
Thus, using the explicit form of $u_k(\tau)$, \eqref{eq:sstt-int1} becomes:
\begin{align}
\Big\langle h_{ij}(\pmb{k}_1) h_{kl}(\pmb{k}_2) H_{(sstt)}(\tau') \Big\rangle
& =
(2\pi)^3 \delta^{(3)}(\pmb{k}_{12}) 
\frac{1}{{\tau'}^2} \frac{H^6}{8(k_1k_2)^3\mpl^6}
\nonumber\\
& \quad
\times
\int \frac{d^3q}{(2\pi)^3} q^2(1-\cos^2\theta) \frac{1}{q^3} \big( 1+q^2{\tau'}^2 \big) \big( 1- ik\tau' \big)^2 e^{2ik\tau'}
\, .
\end{align}
Performing the integration with respect to $q$ up to the cutoff $a(\tau')\Lambda_\text{UV} = -\Lambda_\text{UV}/(H\tau')$, we find:
\begin{equation}
\int_0^{a(\tau')\Lambda_\text{UV}} dq q \big( 1 + q^2{\tau'}^2 \big)
=
\frac{1}{2{\tau'}^2} \bigg( \frac{\Lambda_\text{UV}}{H} \bigg)^2
\Bigg[ 1 + \bigg( \frac{\Lambda_\text{UV}}{H} \bigg)^2 \Bigg]
\, .
\end{equation}
Thus, we find the result is proportional to $\Lambda_\text{UV}^2/H^2$, as stated in the main text.


\begin{thebibliography}{99}




\bibitem{Planck:2018vyg}
N.~Aghanim \textit{et al.} [Planck],
Astron. Astrophys. \textbf{641}, A6 (2020)
[erratum: Astron. Astrophys. \textbf{652}, C4 (2021)]
[arXiv:1807.06209 [astro-ph.CO]].




\bibitem{Guth:1980zm}
A.~H.~Guth,
Phys. Rev. D \textbf{23}, 347-356 (1981).




\bibitem{Linde:1981mu}
A.~D.~Linde,
Phys. Lett. B \textbf{108}, 389-393 (1982).




\bibitem{Albrecht:1982wi}
A.~Albrecht and P.~J.~Steinhardt,
Phys. Rev. Lett. \textbf{48}, 1220-1223 (1982).




\bibitem{Mukhanov:1981xt}
V.~F.~Mukhanov and G.~V.~Chibisov,
JETP Lett. \textbf{33}, 532-535 (1981).




\bibitem{Mukhanov:2005sc}
V.~Mukhanov,
Cambridge University Press, 2005,
ISBN 978-0-521-56398-7




\bibitem{Weinberg:2008zzc}
S.~Weinberg,
Oxford University Press, 2008,
ISBN: 9780198526827





\bibitem{Baumann:2022mni}
D.~Baumann,
Cambridge University Press, 2022,
ISBN 978-1-108-93709-2, 978-1-108-83807-8




\bibitem{Martin:2013tda}
J.~Martin, C.~Ringeval and V.~Vennin,
Phys. Dark Univ. \textbf{5-6}, 75-235 (2014)
[arXiv:1303.3787 [astro-ph.CO]].




\bibitem{Giannantonio:2011ya}
T.~Giannantonio, C.~Porciani, J.~Carron, A.~Amara and A.~Pillepich,
Mon. Not. Roy. Astron. Soc. \textbf{422}, 2854-2877 (2012)
[arXiv:1109.0958 [astro-ph.CO]].




\bibitem{Karagiannis:2018jdt}
D.~Karagiannis, A.~Lazanu, M.~Liguori, A.~Raccanelli, N.~Bartolo and L.~Verde,
Mon. Not. Roy. Astron. Soc. \textbf{478}, no.1, 1341-1376 (2018)
[arXiv:1801.09280 [astro-ph.CO]].




\bibitem{Sohn:2019rlq}
W.~Sohn and J.~Fergusson,
Phys. Rev. D \textbf{100}, no.6, 063536 (2019)
[arXiv:1902.01142 [astro-ph.CO]].




\bibitem{Weinberg:2005vy}
S.~Weinberg,
Phys. Rev. D \textbf{72}, 043514 (2005)
[arXiv:hep-th/0506236 [hep-th]].




\bibitem{Sloth:2006az}
M.~S.~Sloth,
Nucl. Phys. B \textbf{748}, 149-169 (2006)
[arXiv:astro-ph/0604488 [astro-ph]].




\bibitem{Sloth:2006nu}
M.~S.~Sloth,
Nucl. Phys. B \textbf{775}, 78-94 (2007)
[arXiv:hep-th/0612138 [hep-th]].




\bibitem{Seery:2007we}
D.~Seery,
JCAP \textbf{11}, 025 (2007)
[arXiv:0707.3377 [astro-ph]].




\bibitem{Seery:2007wf}
D.~Seery,
JCAP \textbf{02}, 006 (2008)
[arXiv:0707.3378 [astro-ph]].




\bibitem{Dimastrogiovanni:2008af}
E.~Dimastrogiovanni and N.~Bartolo,
JCAP \textbf{11}, 016 (2008)
[arXiv:0807.2790 [astro-ph]].




\bibitem{Adshead:2008gk}
P.~Adshead, R.~Easther and E.~A.~Lim,
Phys. Rev. D \textbf{79}, 063504 (2009)
[arXiv:0809.4008 [hep-th]].




\bibitem{Adshead:2009cb}
P.~Adshead, R.~Easther and E.~A.~Lim,
Phys. Rev. D \textbf{80}, 083521 (2009)
[arXiv:0904.4207 [hep-th]].




\bibitem{Senatore:2009cf}
L.~Senatore and M.~Zaldarriaga,
JHEP \textbf{12}, 008 (2010)
[arXiv:0912.2734 [hep-th]].




\bibitem{Bartolo:2010bu}
N.~Bartolo, E.~Dimastrogiovanni and A.~Vallinotto,
JCAP \textbf{11}, 003 (2010)
[arXiv:1006.0196 [astro-ph.CO]].




\bibitem{Tan:2019czo}
H.~S.~Tan,
JHEP \textbf{10}, 186 (2020)
[arXiv:1907.07706 [gr-qc]].




\bibitem{delRio:2018vrj}
A.~del Rio, R.~Durrer and S.~P.~Patil,
JHEP \textbf{12}, 094 (2018)
[arXiv:1808.09282 [gr-qc]].




\bibitem{Melville:2021lst}
S.~Melville and E.~Pajer,
JHEP \textbf{05}, 249 (2021)
[arXiv:2103.09832 [hep-th]].




\bibitem{LIGOScientific:2016aoc}
B.~P.~Abbott \textit{et al.} [LIGO Scientific and Virgo],
Phys. Rev. Lett. \textbf{116}, no.6, 061102 (2016)
[arXiv:1602.03837 [gr-qc]].




\bibitem{Carr:2020gox}
B.~Carr, K.~Kohri, Y.~Sendouda and J.~Yokoyama,
Rept. Prog. Phys. \textbf{84}, no.11, 116902 (2021)
[arXiv:2002.12778 [astro-ph.CO]].




\bibitem{Carr:2020xqk}
B.~Carr and F.~Kuhnel,
Ann. Rev. Nucl. Part. Sci. \textbf{70}, 355-394 (2020)
[arXiv:2006.02838 [astro-ph.CO]].




\bibitem{Escriva:2022duf}
A.~Escriv\`a, F.~Kuhnel and Y.~Tada,
[arXiv:2211.05767 [astro-ph.CO]].




\bibitem{Harada:2013epa}
T.~Harada, C.~M.~Yoo and K.~Kohri,
Phys. Rev. D \textbf{88}, no.8, 084051 (2013)
[erratum: Phys. Rev. D \textbf{89}, no.2, 029903 (2014)]
[arXiv:1309.4201 [astro-ph.CO]].




\bibitem{Tsamis:2003px}
N.~C.~Tsamis and R.~P.~Woodard,
Phys. Rev. D \textbf{69}, 084005 (2004)
[arXiv:astro-ph/0307463 [astro-ph]].




\bibitem{Kinney:2005vj}
W.~H.~Kinney,
Phys. Rev. D \textbf{72}, 023515 (2005)
[arXiv:gr-qc/0503017 [gr-qc]].




\bibitem{Germani:2017bcs}
C.~Germani and T.~Prokopec,
Phys. Dark Univ. \textbf{18}, 6-10 (2017)
[arXiv:1706.04226 [astro-ph.CO]].




\bibitem{Motohashi:2017kbs}
H.~Motohashi and W.~Hu,
Phys. Rev. D \textbf{96}, no.6, 063503 (2017)
[arXiv:1706.06784 [astro-ph.CO]].




\bibitem{Ballesteros:2017fsr}
G.~Ballesteros and M.~Taoso,
Phys. Rev. D \textbf{97}, no.2, 023501 (2018)
[arXiv:1709.05565 [hep-ph]].




\bibitem{Arnowitt:1962hi}
R.~L.~Arnowitt, S.~Deser and C.~W.~Misner,
Gen. Rel. Grav. \textbf{40}, 1997-2027 (2008)
[arXiv:gr-qc/0405109 [gr-qc]].





\bibitem{Salopek:1990jq}
D.~S.~Salopek and J.~R.~Bond,
Phys. Rev. D \textbf{42}, 3936-3962 (1990)




\bibitem{Maldacena:2002vr}
J.~M.~Maldacena,
JHEP \textbf{05}, 013 (2003)
[arXiv:astro-ph/0210603 [astro-ph]].




\bibitem{Lyth:2004gb}
D.~H.~Lyth, K.~A.~Malik and M.~Sasaki,
JCAP \textbf{05}, 004 (2005)
[arXiv:astro-ph/0411220 [astro-ph]].




\bibitem{Wang:2013zva}
Y.~Wang,
Commun. Theor. Phys. \textbf{62}, 109-166 (2014)
[arXiv:1303.1523 [hep-th]].




\bibitem{Schwinger:1960qe}
J.~S.~Schwinger,
J. Math. Phys. \textbf{2}, 407-432 (1961).




\bibitem{Keldysh:1964ud}
L.~V.~Keldysh,
Zh. Eksp. Teor. Fiz. \textbf{47}, 1515-1527 (1964).




\bibitem{Gong:2001he}
J.~O.~Gong and E.~D.~Stewart,
Phys. Lett. B \textbf{510}, 1-9 (2001)
[arXiv:astro-ph/0101225 [astro-ph]].




\bibitem{Arroja:2008ga}
F.~Arroja and K.~Koyama,
Phys. Rev. D \textbf{77}, 083517 (2008)
[arXiv:0802.1167 [hep-th]].




\bibitem{Huenupi:2024ksc}
J.~Huenupi, E.~Hughes, G.~A.~Palma and S.~Sypsas,
[arXiv:2406.07610 [hep-th]].




\bibitem{Maity:2023qzw}
S.~Maity, H.~V.~Ragavendra, S.~K.~Sethi and L.~Sriramkumar,
JCAP \textbf{05}, 046 (2024)
[arXiv:2307.13636 [astro-ph.CO]].




\bibitem{Kristiano:2022maq}
J.~Kristiano and J.~Yokoyama,
Phys. Rev. Lett. \textbf{132}, no.22, 221003 (2024)
[arXiv:2211.03395 [hep-th]].




\bibitem{Firouzjahi:2023btw}
H.~Firouzjahi,
Phys. Rev. D \textbf{108}, no.4, 043532 (2023)
[arXiv:2305.01527 [astro-ph.CO]].




\bibitem{Ballesteros:2024zdp}
G.~Ballesteros and J.~G.~Egea,
JCAP \textbf{07}, 052 (2024)
[arXiv:2404.07196 [astro-ph.CO]].




\end{thebibliography}
\end{document}